\documentclass[aps,showpacs,preprintnumbers,amsmath,amssymb]{revtex4}
\usepackage{dcolumn}
\usepackage[dvips]{epsfig}

\begin{document}
\baselineskip=0.8 cm
\title{\bf Strong gravitational lensing for the photons coupled to Weyl tensor in a Schwarzschild black hole spacetime}

\author{Songbai Chen\footnote{csb3752@hunnu.edu.cn}, Jiliang Jing
\footnote{jljing@hunnu.edu.cn}}

\affiliation{Institute of Physics and Department of Physics, Hunan
Normal University,  Changsha, Hunan 410081, People's Republic of
China \\ Key Laboratory of Low Dimensional Quantum Structures \\
and Quantum Control of Ministry of Education, Hunan Normal
University, Changsha, Hunan 410081, People's Republic of China\\
Synergetic Innovation Center for Quantum Effects and Applications,
Hunan Normal University, Changsha, Hunan 410081, People's Republic
of China\\State Key Laboratory of Theoretical Physics, Institute of Theoretical Physics, Chinese Academy of Sciences,
Beijing 100190, China}

\begin{abstract}
\baselineskip=0.6 cm
\begin{center}
{\bf Abstract}
\end{center}

We have investigated the strong
gravitational lensing for the photons coupled to Weyl
tensor in a Schwarzschild
black hole spacetime. We find that in the four-dimensional black hole spacetime the equation of motion of the photons depends
not only on the coupling between photon and Weyl tensor, but also on
the polarization direction of the photons. It is quite
different from that in the case of the usual photon without coupling
to Weyl tensor in which the equation of motion is independent of the
polarization of the photon. Moreover, we find that the coupling and the
polarization direction modify the properties of the photon
sphere, the deflection angle, the coefficients in strong field
lensing, and the observational gravitational lensing variables.
Combining with the supermassive central object in our Galaxy, we
estimated three observables in the strong gravitational lensing for
the photons coupled to Weyl tensor.

\end{abstract}

 \maketitle
\newpage
\section{Introduction}

The interaction between the electromagnetic and gravitational fields
should be important in physics because the electromagnetic force and
gravity are two kinds of fundamental forces in nature. In the Lagrangian
of the standard Einstein-Maxwell theory, only the quadratic term of Maxwell tensor is related directly to electromagnetic and gravitational
fields,  which can also be looked as the interaction
between Maxwell field and the spacetime metric tensor. However, the
interactions between electromagnetic field and curvature tensor are
not included in this electromagnetic theory. Drummond \textit{et
al.}\cite{Drummond} found that such kind of the couplings could be
appeared naturally in quantum electrodynamics with the photon
effective action originating from one-loop vacuum polarization on a
curved background spacetime. In this effective field theory, all of
the coupling constants are very small and their values are of the
order of the square of the Compton wave length of the electron
$\lambda_e$ since the coupling between electromagnetic field and
curvature tensor is just a quantum phenomenon in this case. However,
the models with arbitrary coupling constant have been investigated
widely in refs. \cite{Turner,Ni,Solanki,Dereli1,Balakin,Hehl,Bamba}
for some physical motivation. In order to
explain the power-law inflation in the early Universe and the large
scale magnetic fields observed in clusters of galaxies, Turner
\textit{et al} \cite{Turner,Bamba} reconsidered Drummond's model \cite{Drummond}  with the
arbitrary coupling constant and found some interesting effects on the electromagnetic fluctuations.  Ni
\cite{Ni} proposed a classical generalized electromagnetic model in
which electromagnetic field is interacted with curvature tensor
through some special coupled terms. Considering that the coupling
between electromagnetic field and curvature tensor should be emerged
reasonably in the region near the classical supermassive compact
objects at the center of galaxies due to their strong gravity and
high mass density, Ni's model has been investigated widely in
astrophysics \cite{Solanki,Dereli1} and black hole physics
\cite{Balakin,Hehl}. These investigation show that the coupling term
modifies the equations of motion both for the electromagnetic and
gravitational fields, which could yield time delays in the arrival
of gravitational and electromagnetic waves.

Weyl tensor is an important tensor in general relativity, which
describes a type of gravitational distortion in the spacetime. The
couplings between
 Maxwell field and Weyl tensor can be treated actually as a special kind
 of interactions between electromagnetic field and curvature tensors since
 Weyl tensor is a function of Riemann tensor $R_{\mu\nu\rho\sigma}$,
 the Ricci tensor $R_{\mu\nu}$, and the Ricci scalar $R$. The electromagnetic
 theory with Weyl corrections have been investigated extensively in the
 literature \cite{Weyl1,Wu2011,Ma2011,Momeni,Roychowdhury,zhao2013}.
It is shown that the couplings with Weyl tensor change the universal
relation with the $U(1)$ central charge in the holographic conductivity
in the background of anti-de Sitter spacetime \cite{Weyl1} and modify
the properties of the holographic superconductor including the critical
temperature and the order of the phase transition \cite{Wu2011,Ma2011,Momeni,Roychowdhury,zhao2013}.
Moreover, we find that with these couplings the dynamical evolution
and Hawking radiation of electromagnetic field in the black hole
spacetime depend on the coupling parameter and the parity of the field \cite{sb2013}.

 According to general theory of relativity, photons
 would be deviated from their straight path when they
  pass close to a compact and massive body,  and
  the corresponding effects are called as gravitational
   lensing \cite{Einstein,schneider,Darwin}. The images
   of the source stars in the gravitational lensing
   carry the information about the source stars and
   gravitational lens itself, which could help us to identify
 the compact astrophysical objects in the Universe
 and examine further alternative theories of gravity
 in their strong field regime. Many investigations \cite{Einstein,schneider,Darwin,Vir,Vir1,Vir11,Fritt,Bozza1,Eirc1,whisk,Bozza2,Bozza3,Gyulchev,sbnonk,
Bhad1,TSa1,AnAv,gr1,Kraniotis,schen,JH} have studied  the propagation of the free photons in the background spacetimes and probed the effects of the spacetime parameters on the gravitational lensing. In general, gravitational
lensing should be depended both on  the properties of the background spacetime and the behavior of the photon itself.  Eiroa \cite{Eirc2} has studied the behavior of photons in Born-Infeld electrodynamics  and found that in this case photons did not follow geodesics
of the metric but they followed geodesics of an effective metric depending
on the Born-Infeld coupling, which modifies the properties of the gravitational lensing.
Therefore,  it is of interest to probe how the interaction between  photon and spacetime curvature tensor affect the gravitational lensing.
From the previous discussion, we know that the couplings between Maxwell tensor and curvature tensor will change the behavior of electromagnetic field in the background spacetime.
It is well known that light is actually a kind of electromagnetic wave, which means that these couplings will modify the propagation of photons in the spacetime and lead to some particular phenomena of gravitational lensing.
The deflection angle of the photons coupled to Riemann tensor in the weak field limit have been studied in \cite{Drummond}. Since the
weak field limit takes only the first order deviation from
Minkowski spacetime and it is valid only in the region far from the black hole,
it is necessary to investigate further the gravity
lensing in the strong field region near the black hole
because that it starts from complete capture of the coupled photon
and dominates the leading order in the divergence of the deflection
angle. Moreover, in order to probe the universal features of the deflection
angles of the photons coupled to spacetime curvature tensors,
we here will study the strong gravitational lensing in
the Schwarzschild black hole as the photons couple to Weyl tensor, and then explore the effect of these couplings on the deflection angle and
the observables in the strong field limit.

The plan of our paper is organized as follows: In Sec.II,  we derive
the equations of motion for the photons coupled to Weyl
tensor in the four-dimensional static and spherical symmetric
spacetime, which can be obtained from the Maxwell equation with Weyl
corrections by the geometric optics approximation
\cite{Drummond,Daniels,Caip,Cho1,Lorenci}. In Sec.III, we will
study the effects of these coupling on the photon sphere radius
and the deflection angles of light ray in a Schwarzschild black
hole spacetime. In Sec.IV, we will study the physical properties of the strong gravitational
lensing for the coupled photons and then probe the
effects of the coupling constant on the coefficients and the
observables of the gravitational lensing in the strong field limit.
We end the paper with a summary.

\section{Equation of motion for the photons coupled to Weyl tensor}

In this section, we will derive
the equations of motion for the photons coupled to Weyl tensor in the four-dimensional
static and spherical
symmetric spacetime by the geometric optics approximation
\cite{Drummond,Daniels,Caip,Cho1,Lorenci}. We begin with the action of the electromagnetic field coupled to Weyl
tensor in the curved spacetime, which can be expressed as \cite{Weyl1}
\begin{eqnarray}
S=\int d^4x \sqrt{-g}\bigg[\frac{R}{16\pi
G}-\frac{1}{4}\bigg(F_{\mu\nu}F^{\mu\nu}-4\alpha
C^{\mu\nu\rho\sigma}F_{\mu\nu}F_{\rho\sigma}\bigg)\bigg].\label{acts}
\end{eqnarray}
Here $C_{\mu\nu\rho\sigma}$ is the Weyl tensor, which is defined as
\begin{eqnarray}
C_{\mu\nu\rho\sigma}=R_{\mu\nu\rho\sigma}-\frac{2}{n-2}(
g_{\mu[\rho}R_{\sigma]\nu}-g_{\nu[\rho}R_{\sigma]\mu})+\frac{2}{(n-1)(n-2)}R
g_{\mu[\rho}g_{\sigma]\nu}.
\end{eqnarray}
Here $n$ and $g_{\mu\nu}$ are the dimension and metric of the spacetime.
The brackets around indices refers to the antisymmetric part. $F_{\mu\nu}$ is
the usual electromagnetic tensor with a form $F_{\mu\nu}=A_{\nu;\mu}-A_{\mu;\nu}$.
The coupling constant $\alpha$ has
the dimension of length-squared. Varying the action (\ref{acts}) with respect to $A_{\mu}$, one can
obtain easily the corrected Maxwell equation
\begin{eqnarray}
\nabla_{\mu}\bigg(F^{\mu\nu}-4\alpha
C^{\mu\nu\rho\sigma}F_{\rho\sigma}\bigg)=0.\label{WE}
\end{eqnarray}
In order to get the equation of motion of the photons from the above
corrected Maxwell equation (\ref{WE}), one can adopt to the
geometric optics approximation in which the photon wavelength
$\lambda$ is much smaller than a typical curvature scale $L$, but is
larger than the electron Compton wavelength $\lambda_e$, i.e.,
$\lambda_e<\lambda<L$. This ensures that the change of the
background gravitational and electromagnetic fields with the typical
curvature scale can be neglected in the process of photon
propagation \cite{Drummond,Daniels,Caip,Cho1,Lorenci}. With this
approximation, the electromagnetic field strength can be written as
\begin{eqnarray}
F_{\mu\nu}=f_{\mu\nu}e^{i\theta},\label{ef1}
\end{eqnarray}
where $f_{\mu\nu}$ is a slowly varying amplitude and $\theta$ is a rapidly varying phase. This means that the derivative term $f_{\mu\nu;\lambda}$ is not dominated and can be neglected in this approximation. The wave vector is $k_{\mu}=\partial_{\mu}\theta$, which can be treated as the usual photon momentum in the theory of quantum particle. The amplitude $f_{\mu\nu}$
is constrained by the Bianchi identity
\begin{eqnarray}
D_{\lambda} F_{\mu\nu}+D_{\mu} F_{\nu\lambda}+D_{\nu} F_{\lambda\mu}=0,
\end{eqnarray}
which leads to
\begin{eqnarray}
k_{\lambda}f_{\mu\nu}+k_{\mu} f_{\nu\lambda}+k_{\nu} f_{\lambda\mu}=0.
\end{eqnarray}
This means that the amplitude $f_{\mu\nu}$ can be written as
\begin{eqnarray}
f_{\mu\nu}=k_{\mu}a_{\nu}-k_{\nu}a_{\mu},\label{ef2}
\end{eqnarray}
where $a_{\mu}$ is the polarization vector satisfying the condition that
$k_{\mu}a^{\mu}=0$. The amplitude $f_{\mu\nu}$  has just three independent components.
Inserting Eqs.(\ref{ef1}) and (\ref{ef2}) into Eq.(\ref{WE}), one can obtain the equation of motion of photon coupled to Weyl tensor
\begin{eqnarray}
k_{\mu}k^{\mu}a^{\nu}+8\alpha
C^{\mu\nu\rho\sigma}k_{\sigma}k_{\mu}a_{\rho}=0.\label{WE2}
\end{eqnarray}
Obviously, the coupling term with Weyl tensor affects the
propagation of the coupled photon in the background spacetime.

Let us now to consider a four-dimensional static and spherical
symmetric black hole spacetime,
\begin{eqnarray}
ds^2&=&-fdt^2+\frac{1}{f}dr^2+r^2
d\theta^2+r^2\sin^2{\theta}d\phi^2,\label{m1}
\end{eqnarray}
where the metric coefficient $f$ is a function of polar coordinate $r$.
In order to introduce a local set of orthonormal frames, one can use the vierbein fields defined by
\begin{eqnarray}
g_{\mu\nu}=\eta_{ab}e^a_{\mu}e^b_{\nu},
\end{eqnarray}
where $\eta_{ab}$ is the Minkowski metric and the vierbeins
\begin{eqnarray}
e^a_{\mu}=(\sqrt{f},\;\frac{1}{\sqrt{f}},\;r,\;r\sin\theta),
\end{eqnarray}
with the inverse
\begin{eqnarray}
e^{\mu}_a=(\frac{1}{\sqrt{f}},\;\sqrt{f},\;\frac{1}{r},\;\frac{1}{r\sin\theta}).
\end{eqnarray}
Defining the notation for the antisymmetric combination of
vierbeins \cite{Drummond,Daniels}
\begin{eqnarray}
U^{ab}_{\mu\nu}=e^a_{\mu}e^b_{\nu}-e^a_{\nu}e^b_{\mu},
\end{eqnarray}
we find that the complete Weyl
tensor can be rewritten in the following form
\begin{eqnarray}
C_{\mu\nu\rho\sigma}=\mathcal{A}\bigg(2U^{01}_{\mu\nu}U^{01}_{\rho\sigma}-
U^{02}_{\mu\nu}U^{02}_{\rho\sigma}-U^{03}_{\mu\nu}U^{03}_{\rho\sigma}
+U^{12}_{\mu\nu}U^{12}_{\rho\sigma}+U^{13}_{\mu\nu}U^{13}_{\rho\sigma}-
2U^{23}_{\mu\nu}U^{23}_{\rho\sigma}\bigg),
\end{eqnarray}
with
\begin{eqnarray}
\mathcal{A}=-\frac{1}{12r^2}\bigg[r^2f''-2f'r+2f-2\bigg].
\end{eqnarray}
In order to obtain the equation of motion for the coupled photon propagation, one can introduce three linear combinations of momentum components \cite{Drummond,Daniels}
\begin{eqnarray}
l_{\nu}=k^{\mu}U^{01}_{\mu\nu},\;\;\;\;\;\;\;\;\;\;
n_{\nu}=k^{\mu}U^{02}_{\mu\nu},\;\;\;\;\;\;\;\;\;\;
m_{\nu}=k^{\mu}U^{23}_{\mu\nu},\label{pvector}
\end{eqnarray}
together with the dependent combinations
\begin{eqnarray}
&&p_{\nu}=k^{\mu}U^{12}_{\mu\nu}=\frac{1}{k^0}\bigg(k^1n_{\nu}-k^2l_{\nu}\bigg),\nonumber\\
&&r_{\nu}=k^{\mu}U^{03}_{\mu\nu}=\frac{1}{k^2}\bigg(k^0m_{\nu}+k^3l_{\nu}\bigg),\nonumber\\
&&q_{\nu}=k^{\mu}U^{13}_{\mu\nu}=\frac{k^1}{k^0}m_{\nu}+
\frac{k^1k^3}{k^2k^0}n_{\nu}-\frac{k^3}{k^0}l_{\nu}.\label{vect3}
\end{eqnarray}
The vectors $l_{\nu}$, $n_{\nu}$, $m_{\nu}$ are independent and orthogonal to the wave vector $k_{\nu}$.
Using the relationship (\ref{vect3}) and contracting the equation (\ref{WE2}) with $l_{\nu}$, $n_{\nu}$, $m_{\nu}$, respectively, one can find that the equation of motion of the photon coupling with Weyl tensor can be simplified as a set of equations for three independent polarisation components $a\cdot l$, $a\cdot n$, and $a\cdot m$,
\begin{eqnarray}
\bigg(\begin{array}{ccc}
K_{11}&0&0\\
K_{21}&K_{22}&
K_{23}\\
0&0&K_{33}
\end{array}\bigg)
\bigg(\begin{array}{c}
a \cdot l\\
a \cdot n
\\
a \cdot m
\end{array}\bigg)=0,\label{Kk}
\end{eqnarray}
with the coefficients
\begin{eqnarray}
K_{11}&=&(1+16\alpha \mathcal{A})(g_{00}k^0k^0+g_{11}k^1k^1)+(1-8\alpha \mathcal{A})(g_{22}k^2k^2+g_{33}k^3k^3),\nonumber\\
K_{22}&=&(1-8\alpha \mathcal{A})(g_{00}k^0k^0+g_{11}k^1k^1+g_{22}k^2k^2+g_{33}k^3k^3),
\nonumber\\
K_{21}&=&24\alpha \mathcal{A} \sqrt{g_{11}g_{22}}k^1k^2,\;\;\;\;\;\;\;\;\;\;
K_{23}=-24\alpha \mathcal{A}\sqrt{-g_{00}g_{33}}k^0k^3,\nonumber\\
K_{33}&=&(1-8\alpha \mathcal{A})(g_{00}k^0k^0+g_{11}k^1k^1)+(1+16\alpha \mathcal{A})(g_{22}k^2k^2+g_{33}k^3k^3).
\end{eqnarray}
The condition of Eq.(\ref{Kk}) with non-zero solution is $K_{11}K_{22}K_{33}=0$. The first root $K_{11}=0$ leads to the modified light cone
\begin{eqnarray}
(1+16\alpha \mathcal{A})(g_{00}k^0k^0+g_{11}k^1k^1)+(1-8\alpha \mathcal{A})(g_{22}k^2k^2+g_{33}k^3k^3)=0, \label{Kk31}
\end{eqnarray}
which corresponds to the case the polarisation vector $a_{\mu}$ is proportional to $l_{\mu}$ and the strength $f_{\mu\nu}\propto (k_{\mu}l_{\nu}-k_{\nu}l_{\mu})$. The second root $K_{22}=0$ means that $a\cdot l=0$ and $a\cdot m=0$ in Eq.(\ref{Kk}), which implies $a_{\mu}=\lambda k_{\mu}$ and then $f_{\mu\nu}$ vanishes \cite{Drummond}. Thus, this root corresponds to an unphysical polarisation and it should be neglected for general directions of propagation of the coupled photon. The third root is $K_{33}=0$, i.e.,
\begin{eqnarray}
(1-8\alpha \mathcal{A})(g_{00}k^0k^0+g_{11}k^1k^1)+(1+16\alpha \mathcal{A})(g_{22}k^2k^2+g_{33}k^3k^3)=0,\label{Kk32}
\end{eqnarray}
which means that the vector $a_{\mu}=\lambda m_{\mu}$ and the
strength $f_{\mu\nu}=\lambda(k_{\mu}m_{\nu}-k_{\nu}m_{\mu})$.
Therefore, the light-cone condition depends on not only the
coupling between photon and Weyl tensor, but also on the polarizations.
 Moreover, the effects of Weyl tensor on the photon propagation are
 different for the coupled photons with different polarizations,
 which yields a phenomenon of birefringence in the spacetime \cite{Daniels,Caip,Cho1,Lorenci}.
When the coupling
constant $\alpha=0$ the light-cone conditions (\ref{Kk31}) and
(\ref{Kk32}) recover to the usual form without Weyl corrections.

\section{Effects of Weyl corrections on the deflection angles for light ray
in a Schwarzschild black hole spacetime}

In this section, we will study the deflection angles for light ray
as photon couples to Weyl tensor in the background of a
Schwarzschild black hole, and probe the effects of the coupling and
the polarization types on the deflection angle.

For a Schwarzschild black hole spacetime, the metric function is
$f=1-\frac{2M}{r}$ and then the light-cone conditions
(\ref{Kk31}) and
(\ref{Kk32}) can be expressed as
\begin{eqnarray}
(1+\frac{16\alpha M}{r^3})(g_{00}k^0k^0+g_{11}k^1k^1)+(1-\frac{8\alpha M}{r^3} )(g_{22}k^2k^2+g_{33}k^3k^3)=0, \label{Kks31}
\end{eqnarray}
for the photon with the polarization along $l_{\mu}$ (PPL) and
\begin{eqnarray}
(1-\frac{8\alpha M}{r^3})(g_{00}k^0k^0+g_{11}k^1k^1)+(1+\frac{16\alpha M}{r^3} )(g_{22}k^2k^2+g_{33}k^3k^3)=0,\label{Kks32}
\end{eqnarray}
for the photon with the polarization along $m_{\mu}$ (PPM), respectively. The light cone conditions (\ref{Kks31}) and (\ref{Kks32})
imply that the motion of the coupled photons is non-geodesic in the Schwarzschild metric. Actually, these photons follow null geodesics of the effective metric $\gamma_{\mu\nu}$, i.e., $\gamma^{\mu\nu}k_{\mu}k_{\nu}=0$ \cite{Breton}. The effective metric for the coupled photon can be expressed as
\begin{eqnarray}
ds^2=-A(r)dt^2+B(r)dr^2+C(r)W(r)^{-1}(d\theta^2+\sin^2\theta d\phi^2),\label{l01}
\end{eqnarray}
where $A(r)=B(r)^{-1}=1-\frac{2 M}{r}$ and $C(r)=r^2$. The quantity $W(r)$ is
\begin{eqnarray}
W(r)=\frac{r^3-8\alpha M}{r^3+16\alpha M},\label{v11}
\end{eqnarray}
for PPL and
\begin{eqnarray}
W(r)=\frac{r^3+16\alpha M}{r^3-8\alpha M},\label{v12}
\end{eqnarray}
for PPM, respectively.
For simplicity,
we here just consider that both the observer and the source lie in
the equatorial plane in the Schwarzschild black hole spacetime and
the whole trajectory of the photon is limited on the same plane.
With this condition $\theta=\frac{\pi}{2}$, we can obtain the
reduced effective metric in the form
\begin{eqnarray}
ds^2=-A(r)dt^2+B(r)dr^2+C(r)W(r)^{-1}d\phi^2.\label{l1}
\end{eqnarray}
For the photon moving in the equatorial plane ($\theta=\frac{\pi}{2}$), we have $k_2=0$ and $k_{\mu}=(k_0,k_1,0,k_3)$. And then the polarisation vectors $l_{\mu}$ and $m_{\mu}$ can be expressed further as
\begin{eqnarray}
l_{\mu}=(-k^1,k^0,0,0),\;\;\;\;\;\; m_{\mu}=(0,0,-k^3,0).
\end{eqnarray}
This means that $m_{\mu}$ is the polarization orthogonal to the plane
of motion and  $l_{\mu}$ lies on the plane of motion in this case.

In the four dimensional static spacetime (\ref{l01}) with cyclic coordinates $t$ and $\phi$, it is easy to obtain two constants of motion of the geodesics
\begin{eqnarray}
E=-g_{00}\dot{x}^{\mu}=A(r)\dot{t},\;\;\;\;\;\;\;\;\;
L=g_{33}\dot{x}^{\mu}=C(r)W(r)^{-1}\dot{\phi}.
\end{eqnarray}
where a dot represents a derivative with respect to affine parameter
$\lambda$ along the geodesics. $E$ and $L$ correspond to the energy and angular momentum of the coupled photon, respectively. Making use of these two constants and $k^{\mu}=\frac{dx^{\mu}}{d\lambda}$, one can
find that the equations of motion of coupled photon can be expressed further as
\begin{eqnarray}
\bigg(\frac{dr}{d\lambda}\bigg)^2
=\frac{1}{B(r)}\bigg[\frac{E^2}{A(r)}-W(r)\frac{L^2}{C(r)}\bigg].\label{v1}
\end{eqnarray}
Comparing with
Eqs.(8), (16) and (17) in ref.\cite{sb2013}, we find that the
equations of motion (\ref{v1}) correspond actually to the radial
equations of the electromagnetic perturbation with the even parity
and the odd parity in the geometric optics limit, respectively. For
a coupled photon coming from infinite, the form of the deflection
angle in the Schwarzschild black hole spacetime is similar to that
in the case without coupling \cite{Vir1}
\begin{eqnarray}
\alpha(r_{0})=I(r_{0})-\pi,
\end{eqnarray}
where $r_0$ is the closest approach distance. However, $I(r_{0})$
depends on the polarization directions of the photon in this case,
i.e.,
\begin{eqnarray}
I(r_0)=2\int^{\infty}_{r_0}\frac{dr}{\sqrt{A(r) C(r)/ W(r)}
\sqrt{\frac{C(r)A(r_0)W(r_0)}{A(r)W(r)C(r_0)}-1}},\label{int1}
\end{eqnarray}
which means that the deflection angle of
PPL is
different from that of PPM. Obviously, as $\alpha\rightarrow 0$, one can
find that $W(r)\rightarrow 1$ and the deflection angle can be
reduced to that of usual photon \cite{Vir1}.

From Eqs.(\ref{v11}) and (\ref{v12}), we find that there is a
singularity in Eq.(\ref{v1})
 at $r_{sin}=(-16\alpha M)^{1/3}$ for PPL  and
 at $r_{sin}=(8\alpha M)^{1/3}$ for PPM
 because the quantity $\frac{dr}{d\lambda}$ in Eq.(\ref{v1})
 is divergent at this surface.  The position of the singularity depends on the
 coupling parameter and the polarizations of the coupled photon. Considering that a photon should
 propagate continuously in the region outside the event horizon, the coupling constant $\alpha$ must satisfy
 $r^3_H+16\alpha M > 0$ (i.e., $\alpha>\alpha_{c1}=-M^2/2$) for PPL, and satisfy
 $r^3_H-8\alpha M > 0$ (i.e., $\alpha<\alpha_{c2}=M^2$) for PPM.
 With this constraint, the singularity $r_{sin}$ lies inside the event horizon $r_H$ and it does not affect the propagation of the
 photon. And then, we can use the usual methods \cite{Vir,Bozza2} to study the deflection angles for the photon coupled to Weyl tensor in the
background of a Schwarzschild black hole.

Using the photon sphere equation given in \cite{Vir}, one can obtain
that in a Schwarzschild spacetime the impact parameter and the
equation of circular orbits of the coupled photon are
\begin{eqnarray}
&&u=\sqrt{\frac{C(r)}{A(r)W(r)}},\label{u}\\
&&W(r)[A'(r)C(r)-A(r)C'(r)]+A(r)C(r)W'(r)=0.\label{sp}
\end{eqnarray}
Here we set $E=1$. As the coupling parameter
$\alpha\rightarrow 0$, we find that the function $W\rightarrow 1$, which results in
that the impact parameter and the equation of circular photon orbits for PPL are the same as those for PPM. This means that gravitational lensing is
independent of the polarization directions of the photon in the case
without the coupling. Substituting Eqs.(\ref{v11}) and (\ref{v12})
into equation (\ref{sp}), we can obtain the equation of circular photon orbits
\begin{eqnarray}
2(r^3+16\alpha M)(r^3-8\alpha M)(r-3M)\pm72\alpha M r^3(r-2M)=0.\label{sppl}
\end{eqnarray}
Here the signs ``-" and ``+" in the last term correspond to the
cases of PPL and  PPM, respectively. The
biggest real roots of equations (\ref{sppl}) outside the horizon can
be defined as the photon sphere radius of the coupled photons.
\begin{figure}[ht]
\begin{center}
\includegraphics[width=7cm]{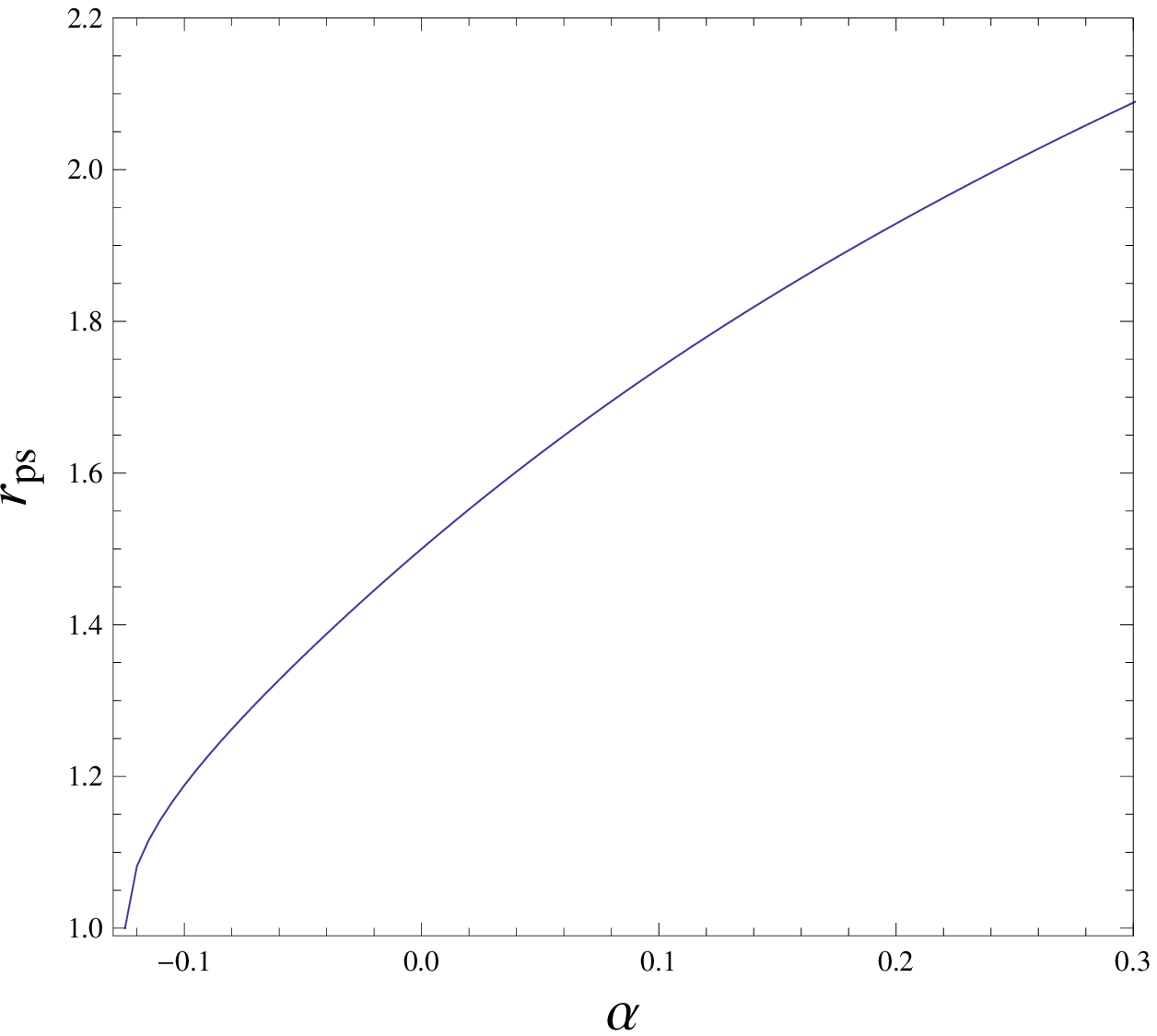}\;\;\;\includegraphics[width=7cm]{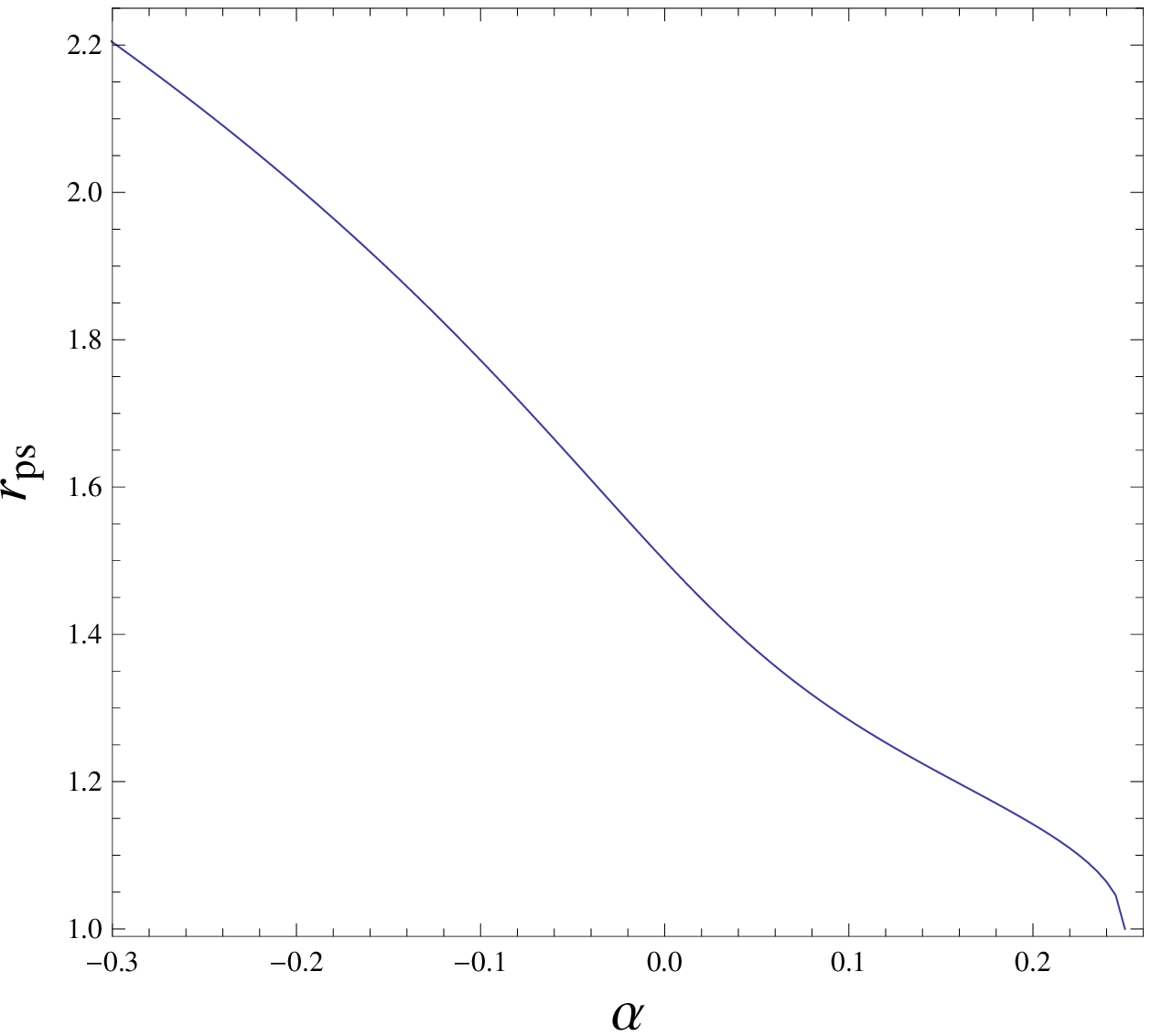}
\caption{Variety of the photon sphere radius $r_{ps}$ with
the coupling constant $\alpha$ in a Schwarzschild black hole
spacetime. The left and the right are for  PPL and PPM,
respectively. Here we set $2M=1$.}
\end{center}
\end{figure}
Obviously, the complex dependence of the equation (\ref{sppl}) on the coupling parameter $\alpha$ yields that we can not get an
analytical form for the photon sphere radius for the
coupled photons. With the help of the numerical method, in Fig.1 we plot the
dependence of the photon sphere radius $r_{ps}$ for the coupled photons  on the coupling parameter $\alpha$. It tells us that with
increase of $\alpha$, $r_{ps}$ increases for PPL and
decreases for PPM in the Schwarzschild black hole
spacetime. This implies that the properties of gravitational lensing
for PPL is different from that for PPM. In
other words, gravitational lensing for the photon coupled to Weyl
tensor depends not only on the coupling constant $\alpha$, but also
on the polarization directions of the photon. Moreover, we also find that as the couple constant $\alpha$ tends to the critical value $\alpha_{c1}$ or $\alpha_{c2}$, the photon sphere of the coupled photon is overlapped with the event horizon of the black hole.

\section{Effects of Weyl Corrections on strong gravitational lensing in a Schwarzschild black hole spacetime}
In this section, we will study the effects of the coupling with Weyl tensor on the coefficients and the
observables of the gravitational lensing in the strong field limit.
Following the evaluation method proposed by Bozza
\cite{Bozza2}, we can define a variable
\begin{eqnarray}
z=1-\frac{r_0}{r},
\end{eqnarray}
and rewrite the integral (\ref{int1}) as
\begin{eqnarray}
I(r_0)=\int^{1}_{0}R(z,r_0)F(z,r_0)dz,\label{in1}
\end{eqnarray}
with
\begin{eqnarray}
R(z,r_0)&=&2\frac{W(r)r^2\sqrt{
C(r_0)}}{r_0C(r)}=2W(z,r_0),
\end{eqnarray}
\begin{eqnarray}
F(z,r_0)&=&\frac{1}{\sqrt{A(r_0)W(r_0)-\frac{A(z,r_0)W(z,r_0)C(r_0)}{C(z,r_0)}}}.
\end{eqnarray}
The function $R(z, r_0)$ is regular for all values of $z
 $ and $r_0$, but the function $F(z, \rho_s)$ diverges as $z$ tends to zero. This is similar to that in the Schwarzschild black hole spacetime without the coupling. Thus, one can split the integral (\ref{in1})  into  the
divergent part $I_D(r_0)$ and the regular one $I_R(r_0)$
\begin{eqnarray}
I_D(r_0)&=&\int^{1}_{0}R(0,r_{ps})F_0(z,r_0)dz, \nonumber\\
I_R(r_0)&=&\int^{1}_{0}[R(z,r_0)F(z,r_0)-R(0,r_{ps})F_0(z,r_0)]dz
\label{intbr},
\end{eqnarray}
where the function $F_0(z,r_{0})$ is obtained by expanding the argument of the square root in $F(z,r_{0})$ to the second order in $z$,
\begin{eqnarray}
F_0(z,r_{0})=\frac{1}{\sqrt{p(r_{0})z+q(r_0)z^2}},
\end{eqnarray}
with
\begin{eqnarray}
p(r_0)&=&-\frac{r_0}{C(r_0)}\bigg\{W(r_0)[A'(r_0)C(r_0)-A(r_0)C'(r_0)]
+A(r_0)C(r_0)W'(r_0)\bigg\},  \nonumber\\
q(r_0)&=&\frac{r_0}{2C(r_0)}\bigg\{2\bigg[C(r_0)-r_0C'(r_0)\bigg]\bigg[A(r_0)W(r_0)C'(r_0)
-C(r_0)\bigg(A(r_0)W(r_0)\bigg)'\bigg] \nonumber\\
&+&r_0C(r_0)\bigg[A(r_0)W(r_0)C(r_0)''
-C(r_0)\bigg(A(r_0)W(r_0)\bigg)''\bigg]\bigg\}.\label{pq}
\end{eqnarray}
If $r_0$ tends to the photon sphere radius $r_{ps}$, one can find that the coefficient $p(r_{0})$ approaches zero and then the integral (\ref{in1}) diverges logarithmically since
the leading term of the divergence in $F(z,r_{0})$ is $z^{-1}$. This means that when the photon is close to the photon sphere, the deflection angle can be approximated as \cite{Bozza2}
\begin{eqnarray}
\alpha(\theta)=-\bar{a}\log{\bigg[\frac{\theta
D_{OL}}{u(r_{ps})}-1\bigg]}+\bar{b}+O[u-u(r_{ps})], \label{alf1}
\end{eqnarray}
with
\begin{eqnarray}
&\bar{a}&=\frac{R(0,r_{ps})}{2\sqrt{q(r_{ps})}},\;\;\;\;\;\;\;\;\; b_R=I_R(r_{ps}), \nonumber\\
&\bar{b}&=
-\pi+b_R+\bar{a}\log{\bigg[\frac{2r^2_{hs}u(r_{ps})''}{u(r_{ps})}\bigg]}.
\label{coa1}
\end{eqnarray}
Here $D_{OL}$ is the distance between observer and gravitational
lens. Substituting Eqs.(\ref{v11}) and (\ref{v12}) into equations (\ref{coa1}), we can obtain the coefficients
$\bar{a}$ and $\bar{b}$ in the strong gravitational lensing formula (\ref{alf1}).
\begin{figure}[ht]
\begin{center}
\includegraphics[width=7cm]{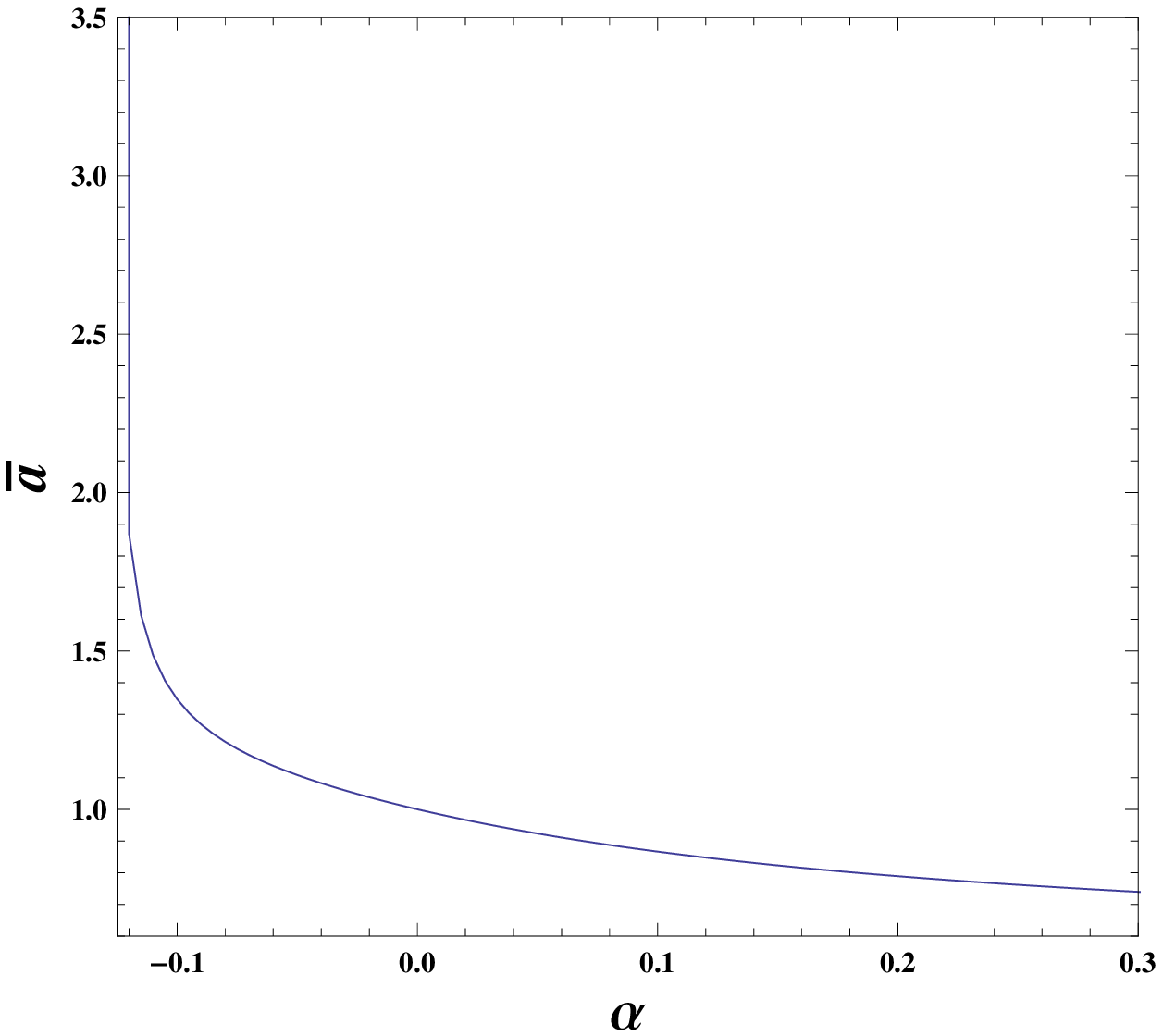}\;\;\;\includegraphics[width=7cm]{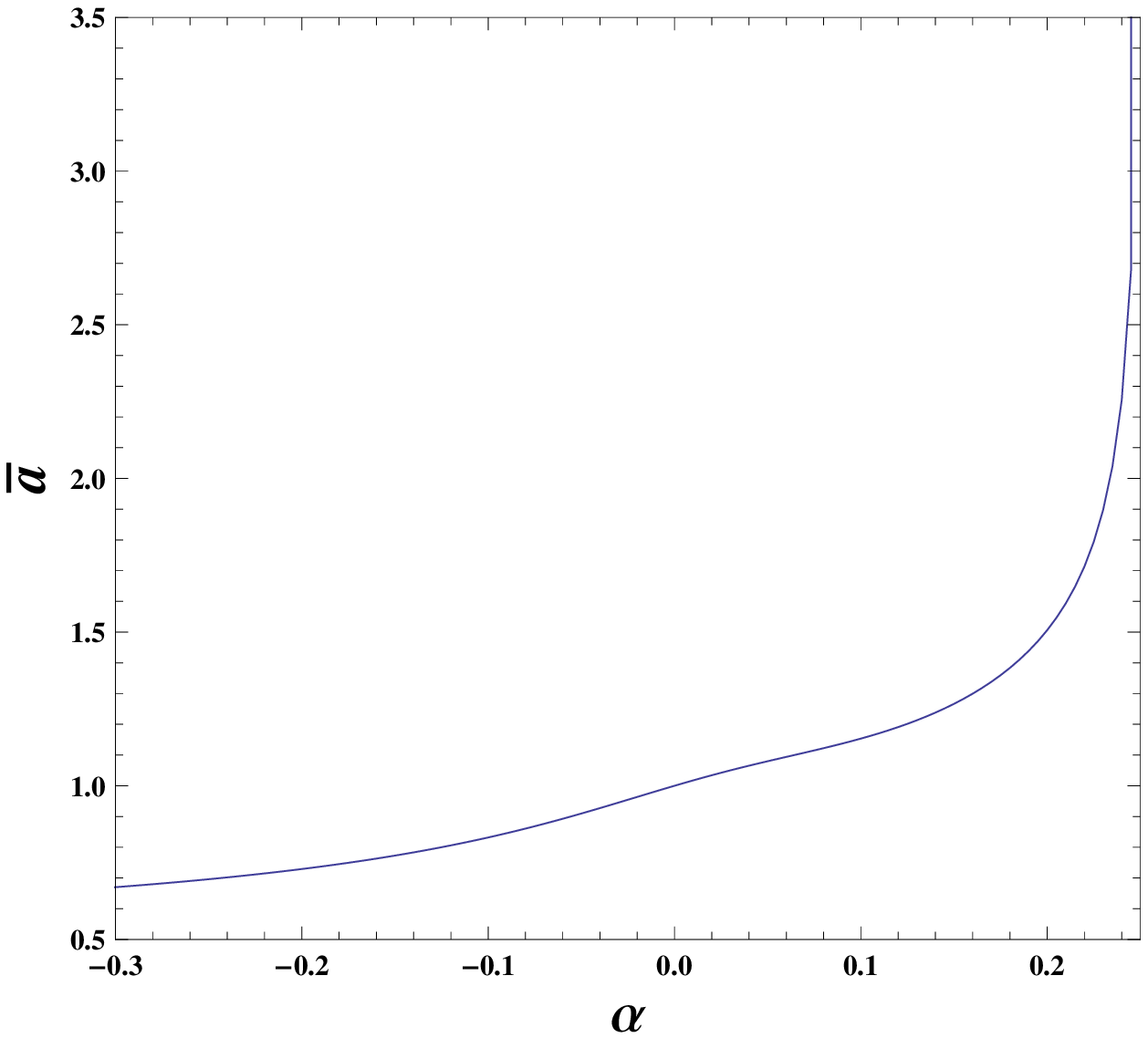}
\caption{Variety of the coefficient $\bar{a}$ with the coupling constant $\alpha$ in a Schwarzschild black hole spacetime. The left and the right are for PPL and PPM, respectively. Here we set $2M=1$.}
\end{center}
\end{figure}
\begin{figure}[ht]
\begin{center}
\includegraphics[width=7cm]{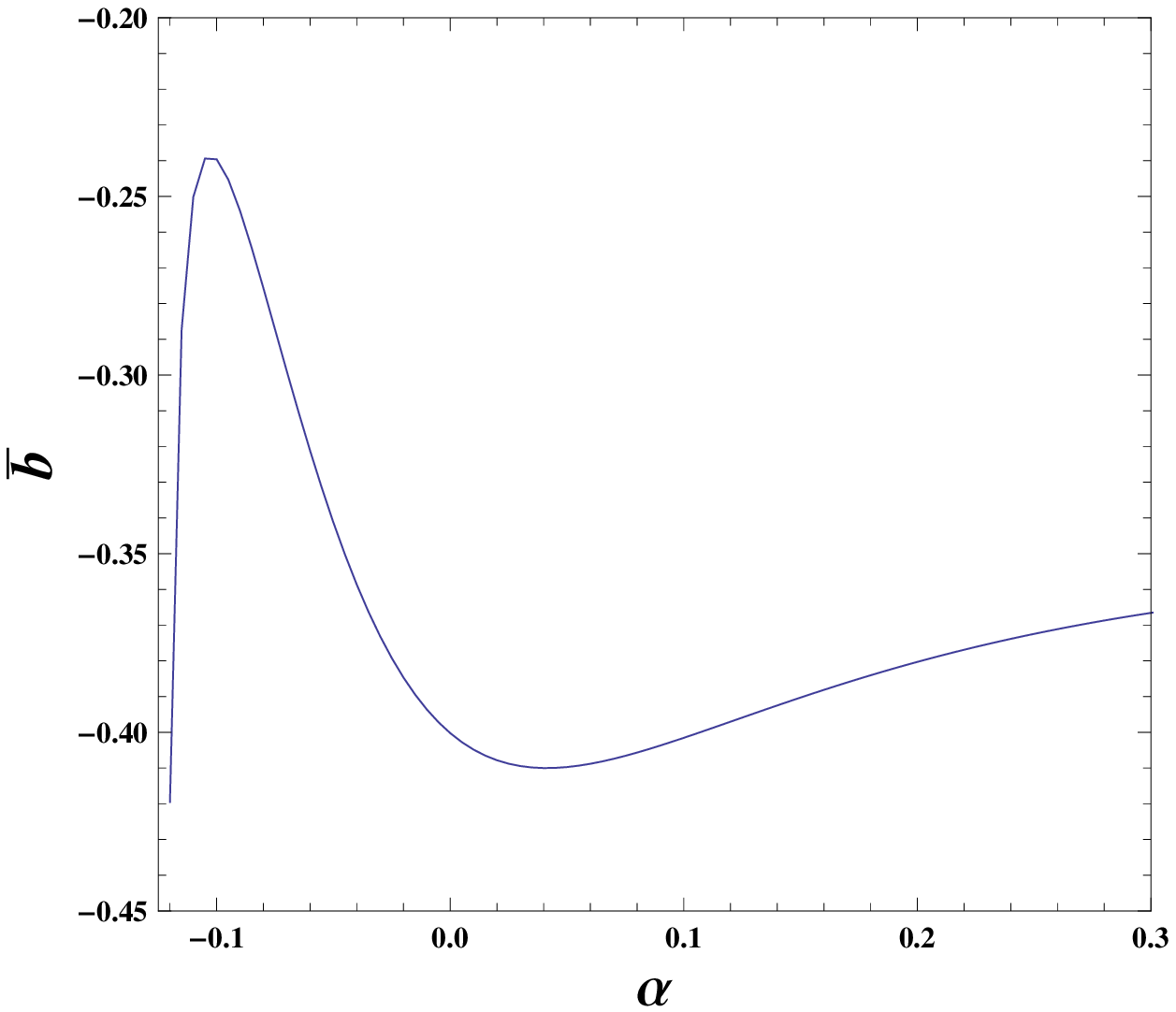}\;\;\;\includegraphics[width=7cm]{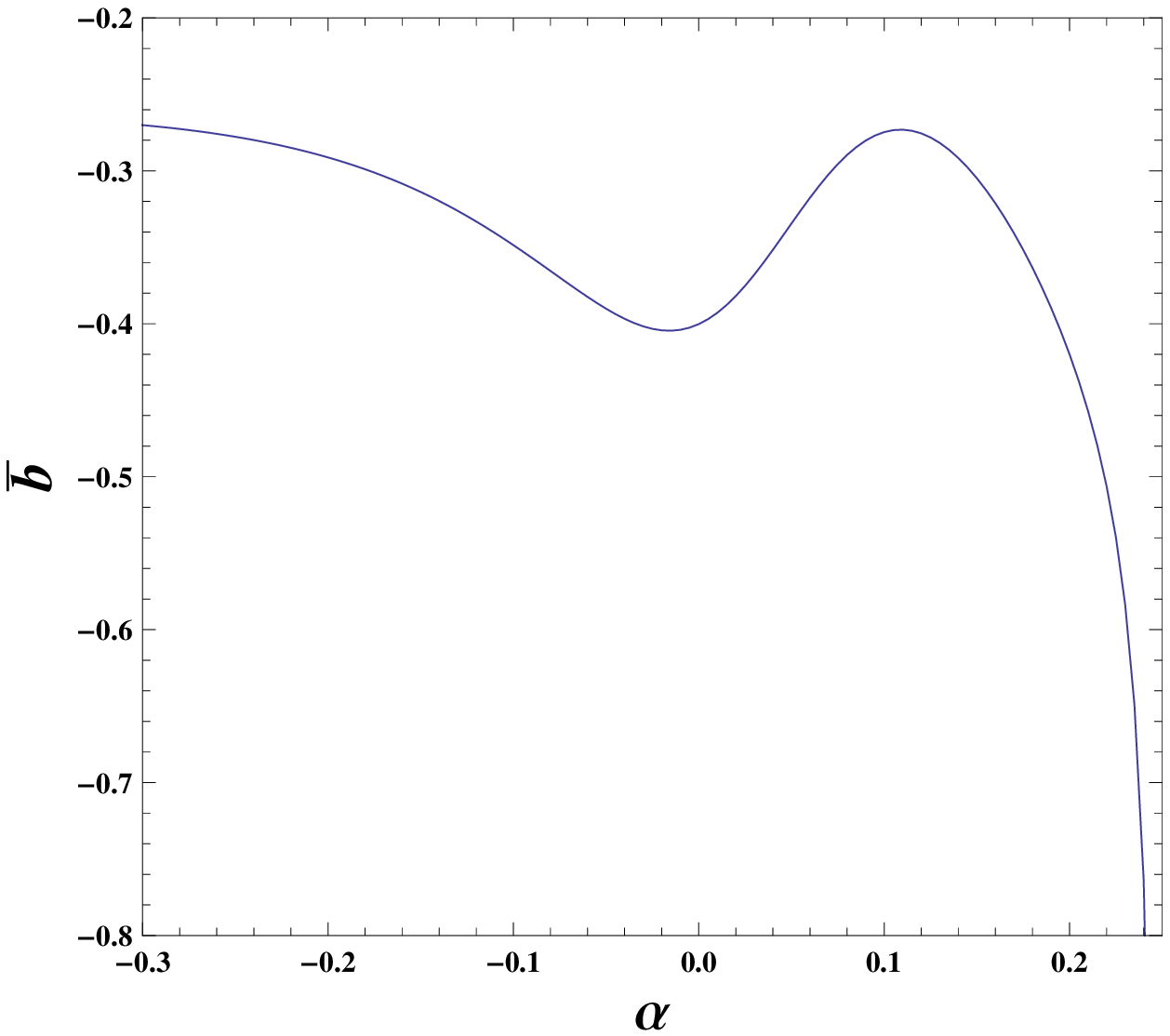}
\caption{Variety of the coefficient $\bar{b}$ with the coupling constant $\alpha$ in a Schwarzschild black hole spacetime. The left and the right are for  PPL and PPM, respectively. Here we set $2M=1$.}
\end{center}
\end{figure}
\begin{figure}[ht]
\begin{center}
\includegraphics[width=7cm]{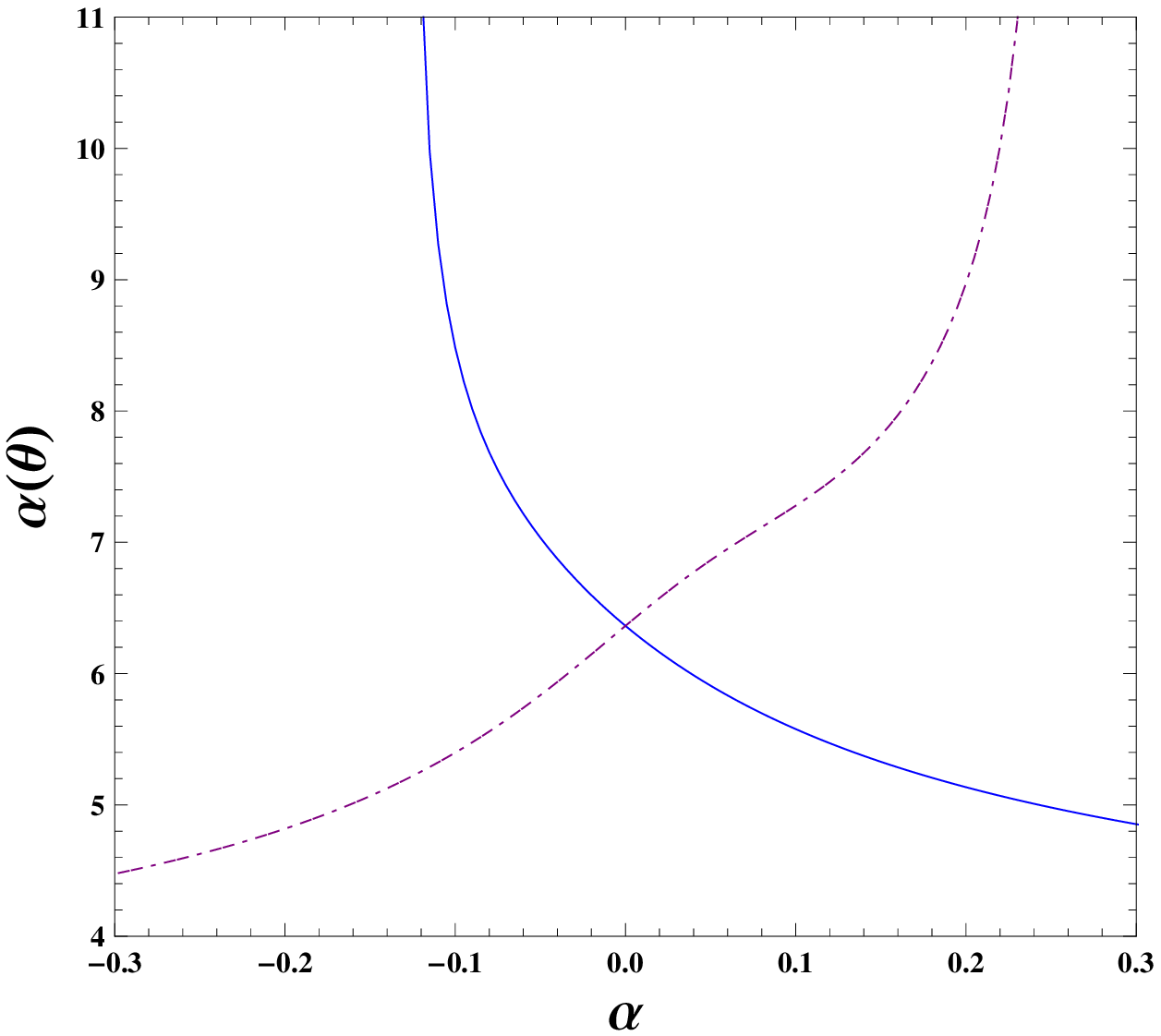}
\caption{Deflection angles in a Schwarzschild
black hole spacetime evaluated at $u=u_{ps}+0.003$ as functions of
$\alpha$. The solid line and the dot-dashed line are for PPL and PPM, respectively. Here we set $2M=1$.}
\end{center}
\end{figure}
In Figs.2-3, we plotted numerically the dependence of the
coefficients ( $\bar{a}$ and $\bar{b}$ ) on the parameter $\alpha$.
It is shown that with the increase of $\alpha$ the coefficient $\bar{a}$ decreases monotonously for PPL and increases for PPM. However, the change of $\bar{b}$ with $\alpha$ is more complicated. For PPL, the coefficient $\bar{b}$ first increases up to its maximum with $\alpha$ and then decreases down to its minimum with the further increase of $\alpha$ ; after that, it increases with $\alpha$ again.
The variety of $\bar{b}$ with $\alpha$ for PPM is converse to that for PPL.
The maximum and minimum of $\bar{b}$ and their corresponding values of $\alpha$ depend on the
 polarization of the coupled photons. Moreover, one can find that the coefficient $\bar{a}$ diverges as the couple constant $\alpha$ tends to the critical value $\alpha_{c1}$ or $\alpha_{c2}$, which implies that the deflection angle in the strong
deflection limit (\ref{alf1}) is not valid in the regime $\alpha<\alpha_{c1}$ for PPL and $\alpha>\alpha_{c2}$ for PPM, which is consistent with the previous discussion. Therefore, the presence of the
coupling makes the change of the coefficients $\bar{a}$ and
$\bar{b}$ more complicated because the effects of the
coupling depend not only on the values of the parameter $\alpha$,
but also on the direction of polarization of the coupled photon.
Furthermore, we plotted in Fig.4 the change of the deflection angles
evaluated at $u=u_{ps}+ 0.003$ with $\alpha$ for PPL  and
PPM, respectively. We find that the deflection angles in
the strong field limit have similar behaviors of the coefficient
$\bar{a}$,
which means that the deflection angles of the light rays are
dominated by the logarithmic term in this case.

We are now in the position to study the effect of the coupling constant $\alpha$
and the direction of polarization of the coupled photon
on the observational gravitational lensing variables in the
strong field limit. If the source and observer are far enough from the
lens, one can find that the lens equation can be simplified further as \cite{Bozza3}
\begin{eqnarray}
\gamma=\frac{D_{OL}+D_{LS}}{D_{LS}}\theta-\alpha(\theta) \; \text{mod}
\;2\pi,
\end{eqnarray}
where $\gamma$ is the angle between the direction
of the source and the optical axis. $D_{LS}$ and $D_{OL}$  are
the lens-source distance and the observer-lens distance, respectively.
 The angle $\theta=u/D_{OL}$ is the angular
separation between the lens and the image. As in
ref.\cite{Bozza3}, we here focus only on the simplest case in which
 the source, lens and observer are
highly aligned so that the angular separation between the lens and
the $n-$th relativistic image can be approximated as
\begin{eqnarray}
\theta_n\simeq\theta^0_n\bigg(1-\frac{u_{ps}e_n(D_{OL}+D_{LS})}{\bar{a}D_{OL}D_{LS}}\bigg),
\end{eqnarray}
with
\begin{eqnarray}
\theta^0_n=\frac{u_{ps}}{D_{OL}}(1+e_n),\;\;\;\;\;\;e_{n}=e^{\frac{\bar{b}+|\gamma|-2\pi
n}{\bar{a}}},\label{st1}
\end{eqnarray}
where $n$ is an integer and $\theta^0_n$ is the image position corresponding
to $\alpha=2n\pi$. In the limit $n\rightarrow \infty$, we have
$e_n\rightarrow 0$, which means that the relationship among the asymptotic position of a
set of images $\theta_{\infty}$, the observer-lens distance $D_{OL}$ and the
minimum impact parameter $u_{ps}$ can be rewritten as a simpler form
\begin{eqnarray}
u_{ps}=D_{OL}\theta_{\infty}.\label{ups}
\end{eqnarray}
In order to get the coefficients $\bar{a}$ and $\bar{b}$, we need
at least another two observations. As in refs.\cite{Bozza2,Bozza3},
we consider a perfect situation in which only the outermost image
 $\theta_1$ is separated as a single image and all the remaining ones
 are packed together at $\theta_{\infty}$.
In this situation  the angular separation $s$ and the relative
magnitudes $r_m$ between the first image and other ones
can be simplified further as \cite{Bozza2,Bozza3}
\begin{eqnarray}
s&=&\theta_1-\theta_{\infty}=
\theta_{\infty}e^{\frac{\bar{b}-2\pi}{\bar{a}}},\nonumber\\
r_m&=&2.5\log{\mathcal{R}_0}=2.5\log{\bigg(\frac{\mu_1}{\sum^{\infty}_{n=2}\mu_n}
\bigg)}
=\frac{5\pi}{\bar{a}}\log{e},\label{sR}
\end{eqnarray}
where $\mathcal{R}_0$ represents the ratio of the flux from the
first image and those from the all other images.
Through measuring these three simple observations $s$, $r_m$,
and $\theta_{\infty}$, one can estimate the coefficients $\bar{a}$, $\bar{b}$
 and the minimum impact parameter $u_{ps}$ in the strong deflection limit.
Comparing their values with those predicted by the coupling theoretical model,
we can extract the characteristics information stored in the strong
gravitational lensing and examine whether this coupling exists in the Universe.
\begin{figure}[ht]
\begin{center}
\includegraphics[width=6cm]{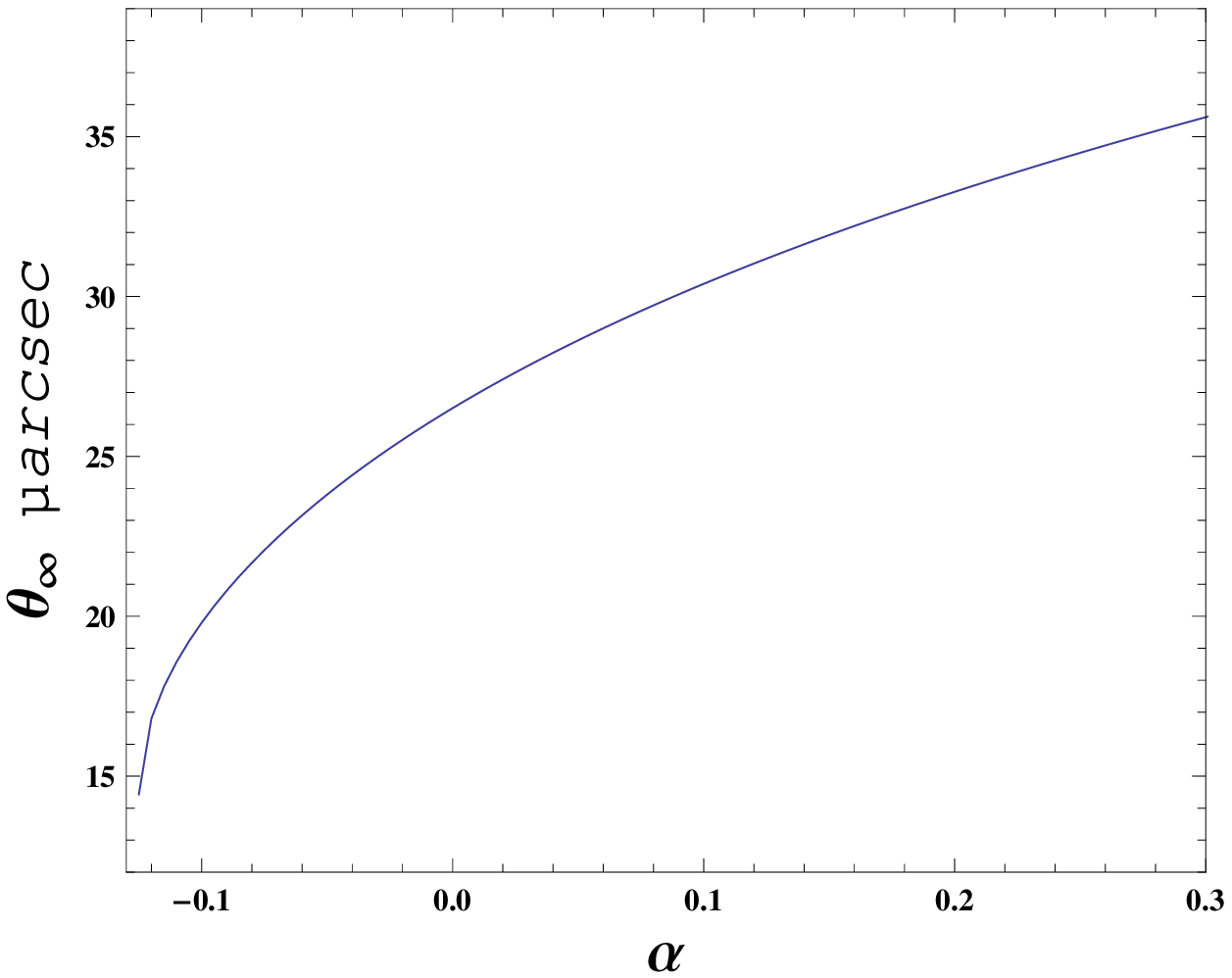}\;\;\;
\includegraphics[width=6cm]{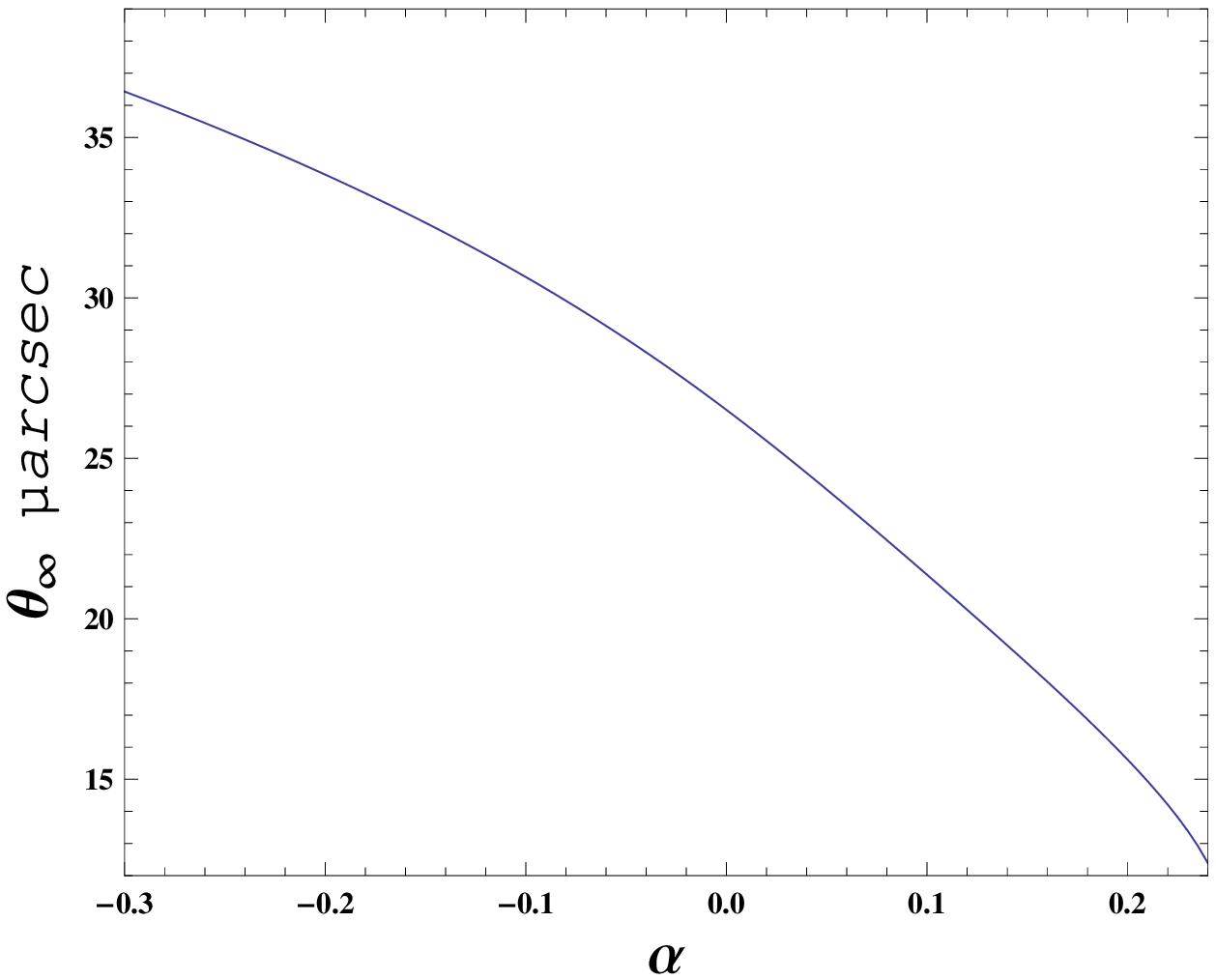}
\caption{Gravitational lensing by the Galactic center black hole.
 The change of  the angular position $\theta_{\infty}$
the coupling constant $\alpha$ in a Schwarzschild black hole spacetime. The left and the right are for PPL and PPM, respectively. Here we set $2M=1$.}
\end{center}
\end{figure}
\begin{figure}[ht]
\begin{center}
\includegraphics[width=6cm]{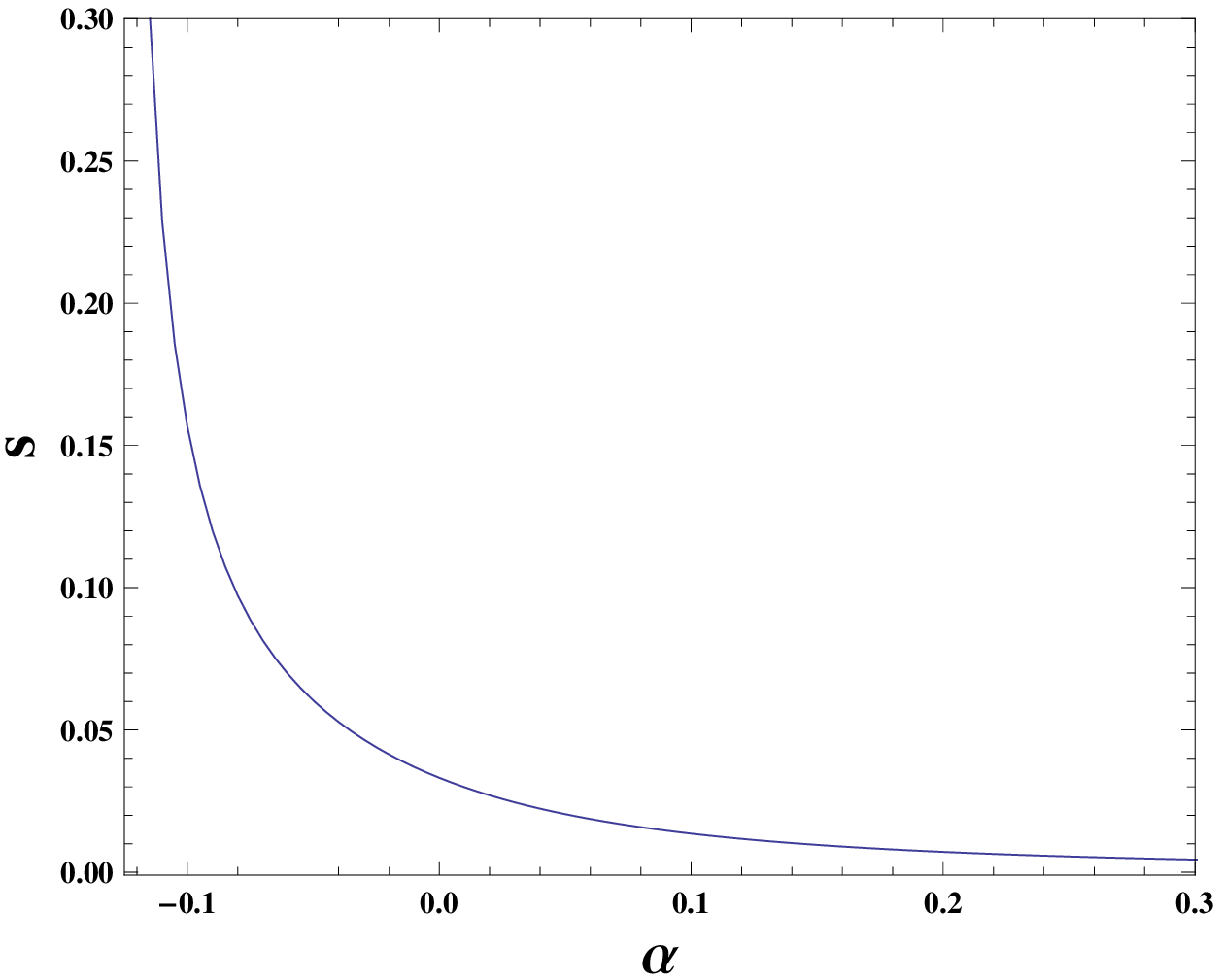}\;\;\;\includegraphics[width=6cm]{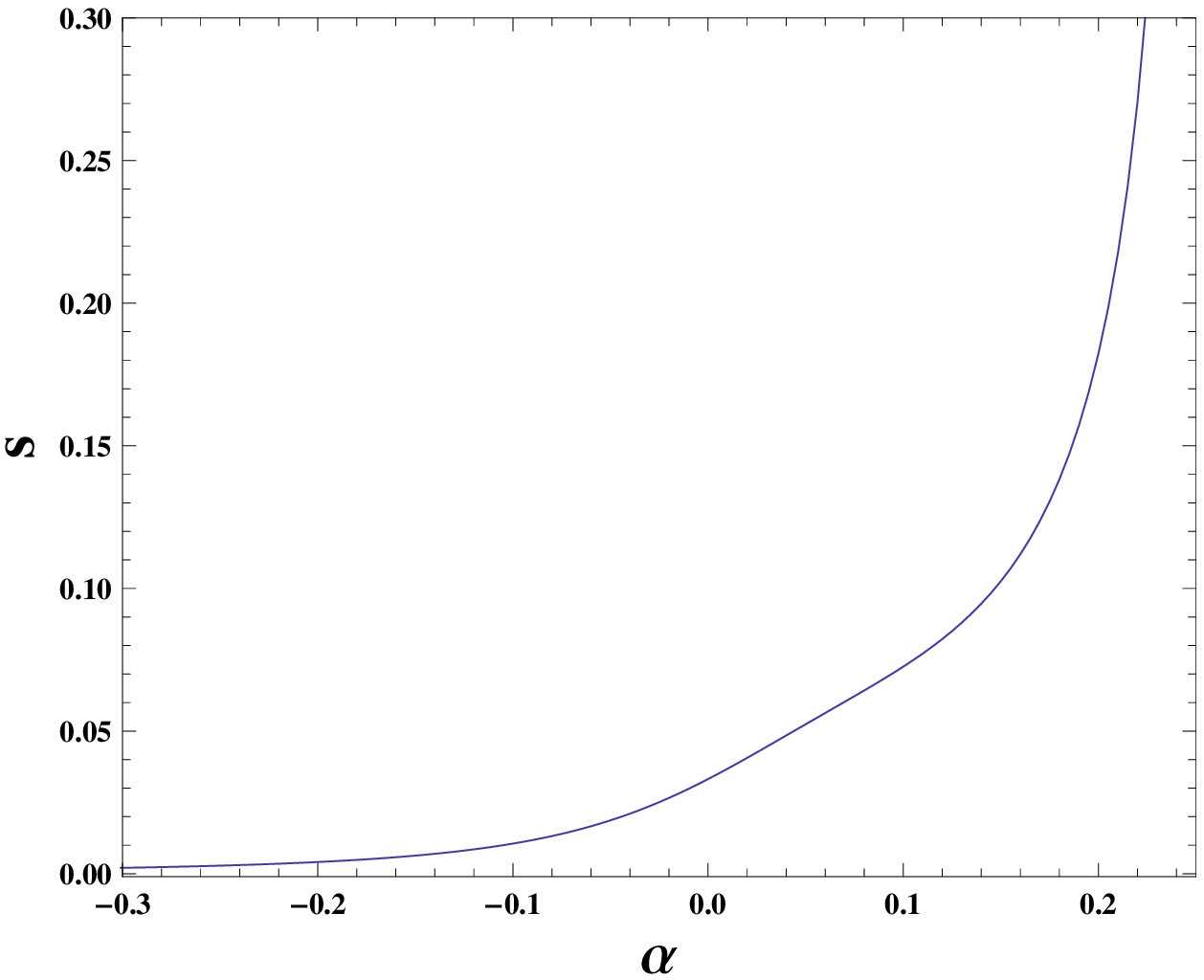}
\caption{Gravitational lensing by the Galactic center black hole.
Variation of the angular separation $s$ with
the coupling constant $\alpha$ in a Schwarzschild black hole spacetime. The left and the right are for  PPL and PPM, respectively. Here we set $2M=1$.}
\end{center}
\end{figure}
\begin{figure}[ht]
\begin{center}
\includegraphics[width=6cm]{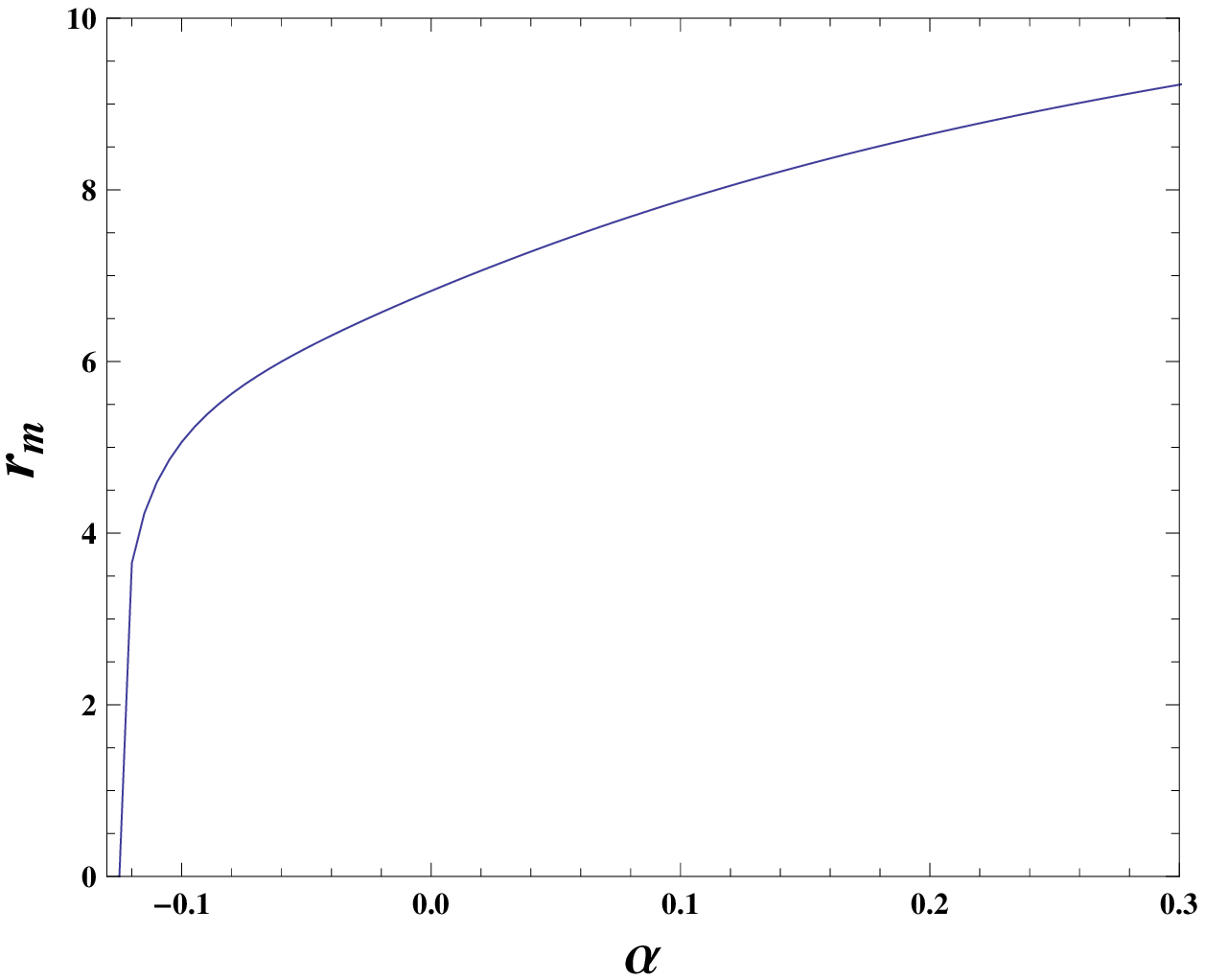}\;\;\;\includegraphics[width=6cm]{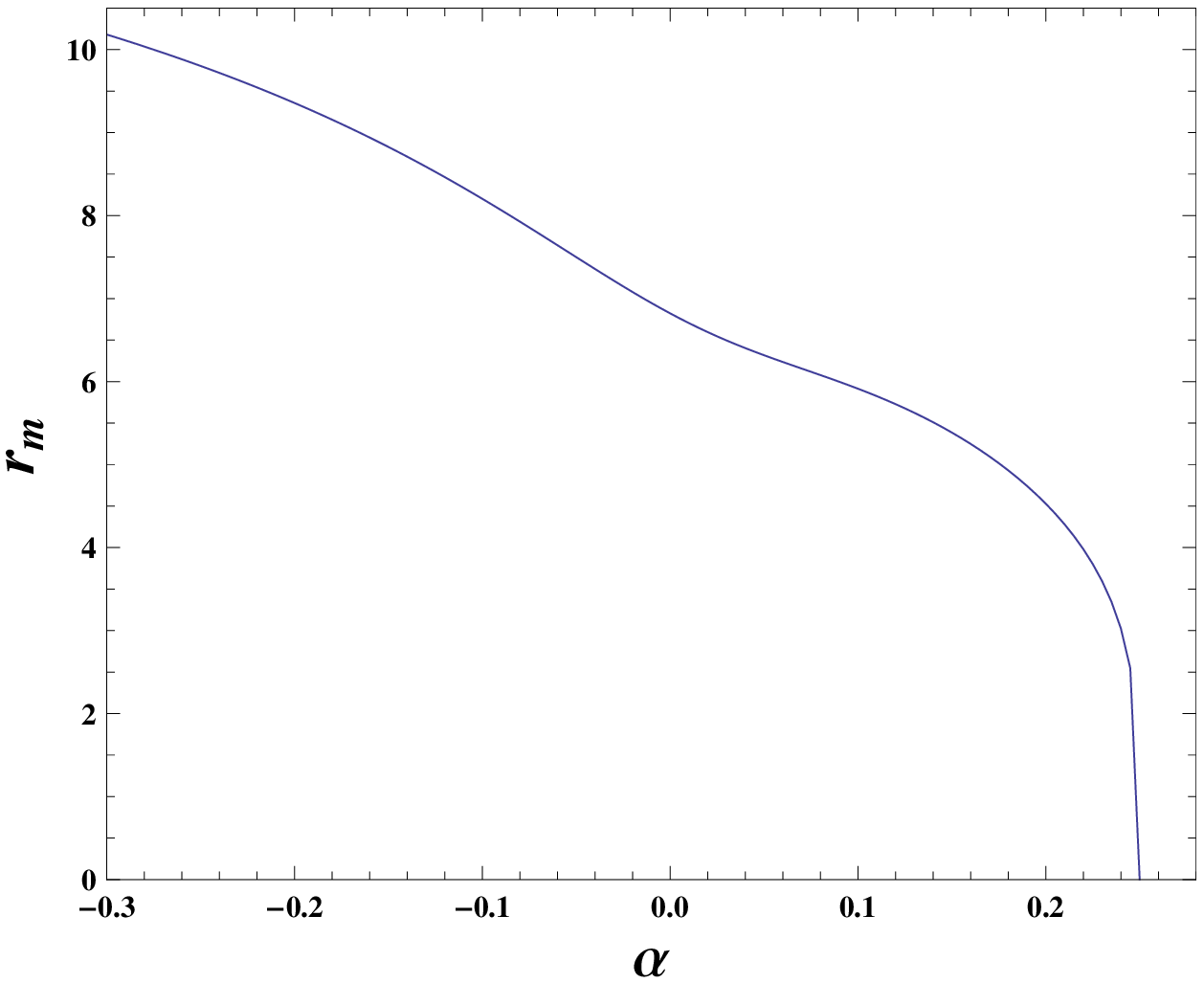}
\caption{Gravitational lensing by the Galactic center black hole.
Variation of the relative magnitudes $r_m$ with
the coupling constant $\alpha$ in a Schwarzschild black hole spacetime. The left and the right are for  PPL and PPM, respectively. Here we set $2M=1$.}
\end{center}
\end{figure}

The mass of the central object of our Galaxy is evaluated to be
$4.4\times 10^6M_{\odot}$ and its distance from the earth is around
$8.5kpc$ \cite{grf}, which means the ratio $GM/D_{OL}
\approx2.4734\times10^{-11}$ . Combing with Eqs. (\ref{coa1}),
(\ref{ups}) and (\ref{sR}), we can estimate the values of the
coefficients and observables in strong gravitational lensing as the
photon couples to Weyl tensor in a Schwarzschild black hole
spacetime. We present the dependence of these observables on the
coupling constant $\alpha$ in Figs.(5)-(7). With the increase of the
coupling constant $\alpha$, both the angular position of the relativistic
images $\theta_\infty$ and the relative magnitudes
$r_m$ increase for PPL and decrease for
PPM.
However, the variety of the angular
separation $s$ with $\alpha$ is converse to the varieties of
$\theta_\infty$ and $r_m$ with $\alpha$. Comparing with those
for the photon without coupling to Weyl tensor, one can find that
 the behaviors of three observables in this case become more complicated.  The main reason is that the coupling between the photon and Weyl tensor changes the equation of motion
of the photon and makes the propagation of the light ray more complicated.

\section{summary}

In this paper, we have investigated the equation of motion of the
photon coupled to Weyl tensor and the corresponding strong
gravitational lensing in a Schwarzschild black hole spacetime. We
find that the coupling constant $\alpha$ and the polarization
direction imprint in the propagation of the coupled photons and
bring some new features for the physical quantities including the
photon sphere radius, the deflection angle, the coefficients ($\bar{a}$ and $\bar{b}$) in the strong
field lensing formulas, and the observational gravitational lensing
variables. With increase of $\alpha$, the photon sphere radius
$r_{ps}$ increases for PPL and decreases for the photon
PPM in the Schwarzschild black hole spacetime.  In the strong
gravity limit,  the coefficient $\bar{a}$ decreases with $\alpha$ for PPL and increases for PPM. The change of $\bar{b}$ with $\alpha$ is more complicated.  Moreover, we find that the coefficient $\bar{a}$ diverges as the couple constant $\alpha$ tends to the critical value $\alpha_{c1}$ or $\alpha_{c2}$.
Combining with the supermassive central object in our
Galaxy, we estimated three observables in the strong gravitational
lensing for the photons coupled to Weyl tensor. It is shown that with the increase of the
coupling constant $\alpha$, both the angular position of the relativistic
images $\theta_\infty$ and the relative magnitudes
$r_m$ increase for PPL and decrease for
PPM.
However, the variety of the angular
separation $s$ with $\alpha$ is converse to the varieties of
$\theta_\infty$ and $r_m$ with $\alpha$ for two different kinds of coupled photons with different polarizations. These indicate that the gravitational lensing
depends not only on the properties of background black hole
spacetime, but also on the polarization of the coupled photon. It
would be of interest to generalize our study to other black hole
spacetimes, such as rotating black holes etc. Work in this direction
will be reported in the future.

\section{\bf Acknowledgments}

This work was  partially supported by the National Natural Science
Foundation of China under Grant No.11275065, the construct
program of key disciplines in Hunan Province and the Open Project Program of State Key Laboratory of Theoretical Physics, Institute of Theoretical Physics, Chinese Academy of Sciences, China (No.Y5KF161CJ1). J. Jing's work was
partially supported by the National Natural Science Foundation of
China under Grant No.11475061.

\end{document}